\documentclass[preprint2]{aastex}

\newcommand{\xmm}{{\em XMM-Newton}}
\newcommand{\chan}{{\em Chandra}}

\newcommand{\psr}{PSR\, J1741$-$2054}

\shorttitle{Optical observations of  \psr}
\shortauthors{Mignani et al.}

\begin{document}

\title{A candidate optical counterpart to the middle-aged $\gamma$-ray pulsar PSR\, J1741$-$2054\thanks{Based on observations collected at the European Organisation for Astronomical Research in the Southern Hemisphere under ESO programme 095.D-0328(B)}}

\author{
R. P. Mignani\altaffilmark{1,2},
V. Testa\altaffilmark{3},
M. Marelli\altaffilmark{1},
A. De Luca\altaffilmark{1,4},
D. Salvetti\altaffilmark{1},
A. Belfiore\altaffilmark{1},
M. Pierbattista\altaffilmark{5},
M. Razzano\altaffilmark{6},
A. Shearer \altaffilmark{7},
P. Moran \altaffilmark{7}
}

\affil{\altaffilmark{1} INAF - Istituto di Astrofisica Spaziale e Fisica Cosmica Milano, via E. Bassini 15, 20133, Milano, Italy}
\affil{\altaffilmark{2} Janusz Gil Institute of Astronomy, University of Zielona G\'ora, Lubuska 2, 65-265, Zielona G\'ora, Poland}
\affil{\altaffilmark{3} INAF - Osservatorio Astronomico di Roma, via Frascati 33, 00040, Monteporzio, Italy }
\affil{\altaffilmark{4} Istituto Nazionale di Fisica Nucleare, Sezione di Pavia, Via Bassi 6, I-27100 Pavia, Italy}
\affil{\altaffilmark{5} Department of Astrophysics and Theory of Gravity, Maria Curie-Sklodowska University, ul. Radziszewskiego 10, 20-031 Lublin, Poland} 
\affil{\altaffilmark{6} Istituto Nazionale di Fisica Nucleare, Sezione di Pisa, I-56127 Pisa, Italy}
\affil{\altaffilmark{7} Centre for Astronomy, National University of Ireland, Newcastle Road, Galway, Ireland}

\begin{abstract}
We carried out deep optical observations of the middle-aged $\gamma$-ray pulsar \psr\ 
with the Very Large Telescope (VLT). 
We identified two objects, of magnitudes $m_v=23.10\pm0.05$ and $m_v=25.32\pm0.08$, at positions consistent with the very accurate \chan\ coordinates of the pulsar,
the faintest of which is more likely to be its counterpart. 
From the VLT images we also detected the known bow-shock nebula around \psr. The nebula is displaced  by $\sim 0\farcs9$ (at the $3\sigma$ confidence level) with respect to its position measured in archival data, showing that the shock propagates in the interstellar medium consistently with the pulsar proper motion. Finally, we could not find evidence of large-scale extended optical emission associated with the pulsar wind nebula detected by \chan, down to a surface brightness limit of $\sim 28.1$ magnitudes arcsec$^{-2}$. Future observations are needed to confirm the optical identification of \psr\ and characterise the spectrum of its counterpart.
  \end{abstract}


\section{Introduction}

\psr\ 
in the Ophiuchus constellation 
is one of the first $\gamma$-ray pulsars to have been discovered  by the Large Area Telescope (LAT; Atwood et al.\ 2009) aboard the {\em Fermi} Gamma-ray Space Telescope through a blind search for pulsations  (Abdo et al.\ 2009).   Its spin period (P$_{\rm s}$= 413 ms) and its derivative ($\dot{P}_{\rm s} = 1.698 \times 10^{-14}$ s s$^{-1}$) yield a characteristic age $\tau = 0.386$ Myr.
The spin parameters also yield a rotational energy loss rate $\dot{E}_{\rm rot} = 9.5 \times 10^{33}$ erg s$^{-1}$ and a dipolar surface magnetic field $B_{\rm s} = 2.68 \times 10^{12}$ G.  Soon after its $\gamma$-ray detection, \psr\ was also discovered as a radio pulsar in archival data from the Parkes telescope. Observations with the Green Bank Telescope yielded a dispersion measure (DM) of 4.7 pc cm$^{-3}$ (Camilo et al.\ 2009) which, from the NE2001 model of the Galactic free electron density  (Cordes \& Lazio 2002), corresponds to a distance of $\sim$ 0.38 kpc. Its very faint radio flux ($\sim 0.16$ mJy at 1.4 GHz; Camilo et al.\ 2009) and  small distance make \psr\ the least luminous radio pulsar.  
\psr\ is also one of the very few middle-aged ($\tau \sim 0.1$--1 Myr)  $\gamma$-ray pulsars closer than $\sim$ 0.5 kpc (see Abdo et al.\ 2013).

The X-ray counterpart to \psr\ was found by {\em Swift} (Camilo et al.\ 2009) soon after its detection as a radio pulsar. The X-ray identification was confirmed by  \chan\ observations that also detected a compact pulsar-wind nebula (PWN) around \psr\ and an $\sim 1\farcm5$-long  trail of X-ray emission (Romani et al.\ 2010). The pulsar was also observed with \xmm\ (Marelli et al.\  2014) and X-ray pulsations 
  were detected for the first time.  Its X-ray spectrum is characterised by the combination of a power-law (PL) and a black body (BB), like in other middle-aged pulsars,  produced by emission from the neutron star magnetosphere and the cooling neutron star surface. An analysis  of the \chan\ data was done by Karpova et al.\ (2014) and Auchettll et al.\  (2015), who also measured the pulsar proper motion ($\mu = 109\pm10$ mas yr$^{-1}$) from the analysis of multi-epoch observations. The proper motion corresponds to a transverse velocity of $196\pm18$ km s$^{-1}$ for a pulsar distance of 0.38 kpc.

The close distance to \psr\ makes it a promising target for a detection in the optical band. Indeed, the middle-aged $\gamma$-ray pulsars PSR\, B0656+14, Geminga, and PSR\, B1055$-$52 have been all identified in the optical (Abdo et al.\ 2013 and references therein) thanks to their distances,  closer than $\sim 0.5$ kpc.  Observations of the \psr\ field in the H$_{\alpha}$ band (Romani et al.\ 2010; Brownsberger \& Romani 2014) allowed to discover a bow-shock nebula, produced by the pulsar motion in the interstellar medium (ISM), whose axis of symmetry is aligned with the major axis of the PWN trail and with the direction of the pulsar proper motion (Auchettl et al.\ 2015).  The pulsar, however, remained undetected at optical wavelengths.   Interestingly, optical spectroscopy of the bow-shock nebula (Romani et al.\ 2010) yielded a pulsar space velocity consistent with that inferred from its proper motion and the DM distance of 0.38 kpc, suggesting that this value is qualitatively correct.

Here, we present the analysis of deep optical observations of \psr\ carried out with the ESO Very Large Telescope (VLT). Observations and data analysis are described in Sectn.\ 2, while the results are presented and discussed in Sectn.\ 3 and 4, respectively.

\section{Observations and Data Reduction}

\psr\ was observed in service mode on May 14 and 15, 2015 with the VLT at the ESO Paranal Observatory
and the second FOcal Reducer and low dispersion Spectrograph  (FORS2;
Appenzeller  et  al.\ 1998)
in imaging mode. The camera was equipped with its default MIT detector, a mosaic of two 4k$\times$2k CCDs aligned along the long axis, optimised  for wavelengths  longer  than 6000  \AA.   With the FORS2 high-resolution collimator,   the  detector  has  a  pixel   scale  of  0\farcs125 (2$\times$2 binning)
and a projected field--of--view (FOV) of 4$\farcm15  \times 4\farcm15$.
The observations were executed with the standard low-gain and fast read-out mode
and through the high-throughput  $b_{\rm HIGH}$ ($\lambda=4400$ \AA;  $\Delta \lambda=1035$\AA)  and $v_{\rm HIGH}$ ($\lambda=5570$ \AA;  $\Delta \lambda=1235$\AA) filters.  To allow for  cosmic-ray removal and reduce the impact of bright star saturation, we obtained  sequences of 30 short exposures (180 s) for a total integration time of 5400 s in both the  $b_{\rm HIGH}$ and $v_{\rm HIGH}$ filters.   Exposures  were taken  in  dark time  and under  clear sky conditions, with the target close to the zenith (airmass $\la 1.1$) and seeing $\sim 0\farcs4$.

We reduced the data (bias  subtraction and  flat--fielding) using tools in the {\sc IRAF}\footnote{IRAF is distributed by the National Optical Astronomy Observatories, which are operated by the Association of Universities for Research in Astronomy, Inc., under cooperative agreement with the National Science Foundation.} package {\sc ccdred}. Per each band, we aligned and average-stacked the reduced  science images with the  {\tt  drizzle} task in {\sc IRAF}, applying a $\sigma$ clipping to filter  out hot/cold pixels and cosmic ray hits.  
We applied the  photometric calibration by using  the FORS2 night  zero points and the atmospheric extinction coefficients\footnote{\texttt{www.eso.org/observing/dfo/quality/FORS2/qc/qc1.html}}.  
We computed the astrometry calibration using the {\em wcstools}\footnote{\texttt{http://tdc-www.harvard.edu/wcstools}} suite of programs and reference stars from the GSC2.3 (Lasker et al.\ 2008).  We obtained mean residuals of $\la 0\farcs1$ in the radial direction, using 30 non-saturated GSC2.3 stars selected to avoid the vignetted regions of the detector. Thanks to the pixel scale of the FORS2 images (0\farcs125), the uncertainty on the centroids of the reference stars is negligible.  To this value we added in quadrature the uncertainty of the image registration  on the GSC-2.3 reference frame  ($\sim$ 0\farcs11) and  the 0\farcs15  uncertainty on the link of the GSC2.3 to the International Celestial Reference Frame. We ended up with an overall accuracy of $\sim$0\farcs2 on our absolute astrometry.

\section{Results}

As a reference to search for the \psr\  optical counterpart, we used its \chan\ coordinates $\alpha =17^{\rm h}  41^{\rm m} 57\fs28$; $\delta  = -20^\circ 54\arcmin 11\farcs8$ (MJD 55337), with an estimated accuracy of 0\farcs3 (Romani et al.\ 2010). 
We note that {\em SIMBAD} currently reports the pulsar $\gamma$-ray coordinates
from the Third {\em Fermi}-LAT $\gamma$-ray source catalogue (Acero et al.\ 2015), whereas those in  the ATNF pulsar data base (Manchester et al.\ 2005) are inconsistent with the published ones (Camilo et al.\ 2009; Romani et al.\ 2010).  We accounted for the \chan\ proper motion\footnote{Note that $\mu_{\delta}$ is reported with the wrong sign at page 70 of  Auchettll et al.\ (2015).} 
$\mu_{\alpha} cos(\delta)=-63\pm12$ mas yr$^{-1}$ and $\mu_{\delta} = -89\pm9$ mas yr$^{-1}$  (Auchettl et al.\ 2015) to extrapolate the pulsar coordinates at the epoch of our VLT observations (MJD 57156). The error on the proper motion produces an uncertainty on the coordinate extrapolation ($\sim 0\farcs06$) negligible compared to that of the reference \chan\ coordinates and the accuracy of our astrometry calibration.

\begin{figure}
\vspace{0.5cm}
\begin{center}
\includegraphics[width=8cm,height=8cm, bb= 0 0 382 382, clip=]{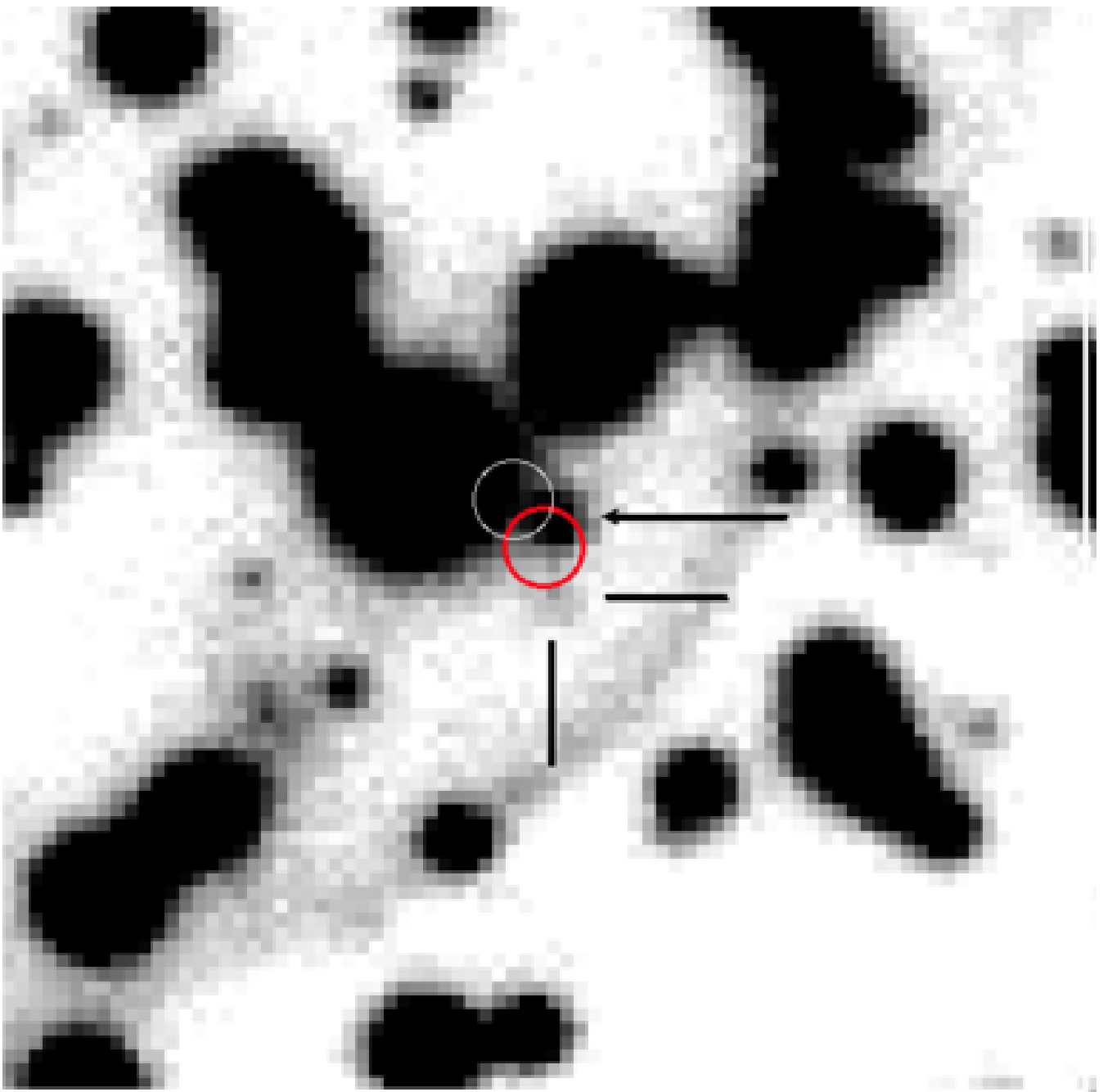}
\includegraphics[width=8cm,height=8cm, bb=0 0 476 476, clip=]{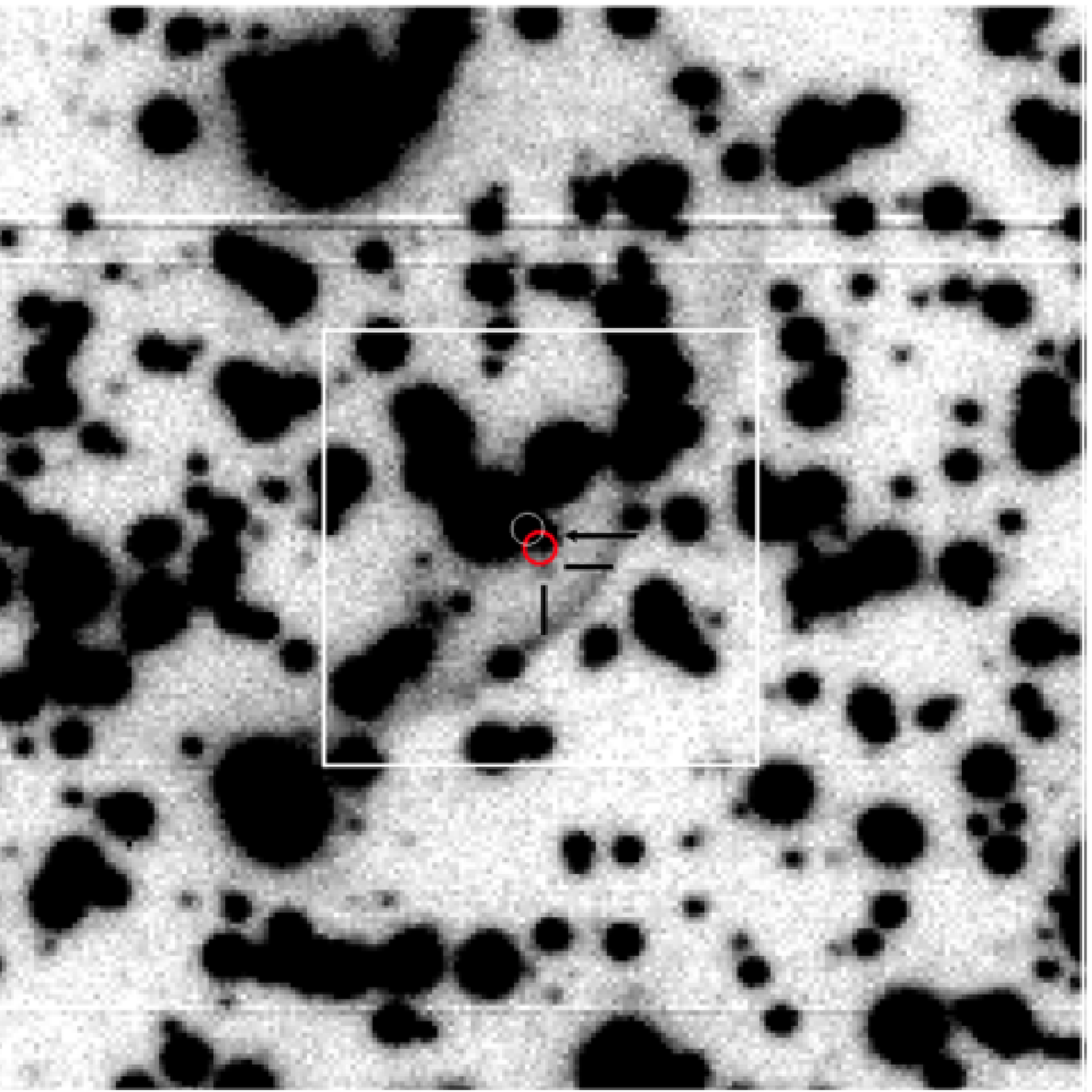}
\caption{{\em Left}: FORS2 image (10\arcsec$\times$10\arcsec) of the \psr\ field (b$_{\rm HIGH}$ filter). North to the top, East to the left. The circles (0\farcs36 radii, accounting for  our astrometry accuracy) mark the reference pulsar position (Romani et al.\ 2010; MJD 55337; white) and that corrected for the proper motion (Auchettl et al.\ 2015) at the VLT epoch (MJD 57156; red).  
Two objects, marked by the arrow and the two ticks, are visible near to it. {\em Right}:  Wider zoom (25\arcsec$\times$25\arcsec) of the same area with the image contrast adjusted to better show the arc-like structure south and west of the pulsar.  \label{fc}}
\end{center}
\end{figure}

A close-up of the FORS2 b$_{\rm HIGH}$-band image centred on the extrapolated \psr\ position is shown in Fig. \ref{fc} (left). An object is clearly detected  within the position error circle, of magnitudes 
$m_b = 24.76\pm0.07$ and $m_v=23.10\pm0.05$.
 A second, fainter object ($m_b = 26.45\pm0.10$ and $m_v=25.32\pm0.08$) is visible south of it
  within $\sim 1.5 \sigma$ 
  from the expected pulsar position.  To minimise the effects of the bright stars north of the pulsar position, we computed the objects magnitudes through PSF photometry with the {\textsc DAOPHOT II} package (Stetson 1994) in {\sc IRAF} and applied the aperture correction.
Owing to its position close to the Galactic plane ($l=6.4^{\circ}$; $b=4.9^{\circ}$),  the field of \psr\ is very crowded. Therefore, we cannot rule out the possibility of
a chance coincidence with the proper motion-corrected \chan\ position.
 Assuming a Poisson distribution, the probability of a having at least a spurious association within a given matching radius
is $P=1-\exp(-\lambda)$, where  $\lambda = \pi\rho r^2$,  $r$ is the association radius 
and $\rho$ is  the number density of field objects of brightness 
between that of the two objects and the image detection limit.
For  $\rho\sim 0.14$ arcsec$^{-2}$, as measured by the number of objects counted in the FORS2 images, and $r \sim 0\farcs46$, chosen as the angular separation between the centre of the error circle and the southernmost of the two objects, we derived 
that $P\sim 0.09$. If \psr\ had an optical luminosity of the same order of magnitude as PSR\, B0656+14, Geminga, and PSR\, B1055$-$52, which are at a similar  distance, we would expect that its optical brightness be in the range $m_v\approx25$--26.   This would make the faintest of the two objects above a more likely candidate counterpart to the pulsar.  Without further evidence, we cannot firmly rule out the other object as a candidate counterpart, though. In this case, however, \psr\ would be about ten times brighter than the other three middle-aged pulsars, unless its distance is overestimated by a factor of three. 
Such a small distance would be incompatible with the 
significant hydrogen column density  $N_{\rm H}$ in the pulsar direction (Marelli et al.\ 2014) and with the measurements of the pulsar space velocity (Romani et al.\ 2010; Auchettl et al.\ 2015), which are in agreement with the DM-based distance.
Further observations will solve this possible ambiguity, e.g. by measuring for the candidate counterpart the same proper motion as  the pulsar,  which would secure its identification.
Fig. \ref{fc} (right) shows a zoom out of the FORS2 image around the \psr\ position.  A region of  extended emission with an arc-like structure is clearly visible around the pulsar, with a spatial extent and morphology very similar to those of the bow-shock nebula detected in H$_{\alpha}$ by Romani et al.\ (2010).  The arc-like structure in Fig. \ref{fc}  is not seen in the v$_{\rm HIGH}$-band image.  We identify this structure with the bow-shock nebula around the pulsar
and attribute its origin to the contribution of the H$_{\gamma}$ and H$_{\beta}$ emission lines in the bow-shock spectrum 
(Romani et al.\ 2010). The wavelengths of these lines fall within the band width of the b$_{\rm HIGH}$ filter but not of the v$_{\rm HIGH}$ one 
(see Sectn.\ 2),
which explains the non-detection of the structure in the latter filter. 

We compared both the position and morphology of the bow-shock nebula with those observed in images taken back on August 21 2009 (MJD=55063)  by Romani et al.\ (2010)
with the ESO New Technology Telescope (NTT).
The data set, retrieved from the public ESO archive, includes
 three exposure in H$_{\alpha}$ (600 s each)
taken with the ESO Faint Object Spectrograph and Camera (EFOSC; Buzzoni et al.\ 1984), with a spatial resolution of 0\farcs25/pixel. We reduced the data and applied the astrometry calibration as described in Sectn.\ 2.  Fig.\  2 (left) shows the 
NTT image with the contours of the 
VLT/FORS2 image overlaid.  Although the morphology of the bow-shock nebula is consistent in the two images, its position in the VLT one is slightly offset to the southwest, i.e. close to the direction of the pulsar proper motion (position angle $215^{\circ}\pm6^{\circ}$, measured east of north). We measured this displacement by comparing the relative positions of the peaks of the nebula surface brightness measured in rectangle of 3\farcs5$\times$1\farcs6 
around its axis of symmetry 
and found that it amounts to $0.94^{+0.20}_{-0.31}$ arcsec. The time span between
the VLT and NTT images ($\sim$ 5.7 years) implies an annual nebula displacement of  $169^{+35}_{-54}$ mas yr$^{-1}$, consistent 
with the pulsar proper motion ($109\pm10$ mas yr$^{-1}$). Therefore, the nebula displacement is 
explained by the shock propagation in the ISM, as a result of the pulsar proper motion.  We also note that for the northern rim of the nebula there is a hint of a larger displacement than for the southern one. This suggests a different propagation velocity of the shock along the north-south direction, probably 
due to a difference in the local density of the ISM.   PSR\, J1741$-$2054 is, thus,   the third pulsar only for which a displacement of the  $H_{\alpha}$ bow-shock nebula has been detected after PSR\, B2224+65 with its renown ``Guitar'' nebula (Chatterjee \& Cordes 2002), and PSR\, J0437$-$4715 (e.g., Brownsberger et al.\ 2015).

\begin{figure}
\vspace{0.5cm}
\begin{center}
\includegraphics[height=8cm,clip=]{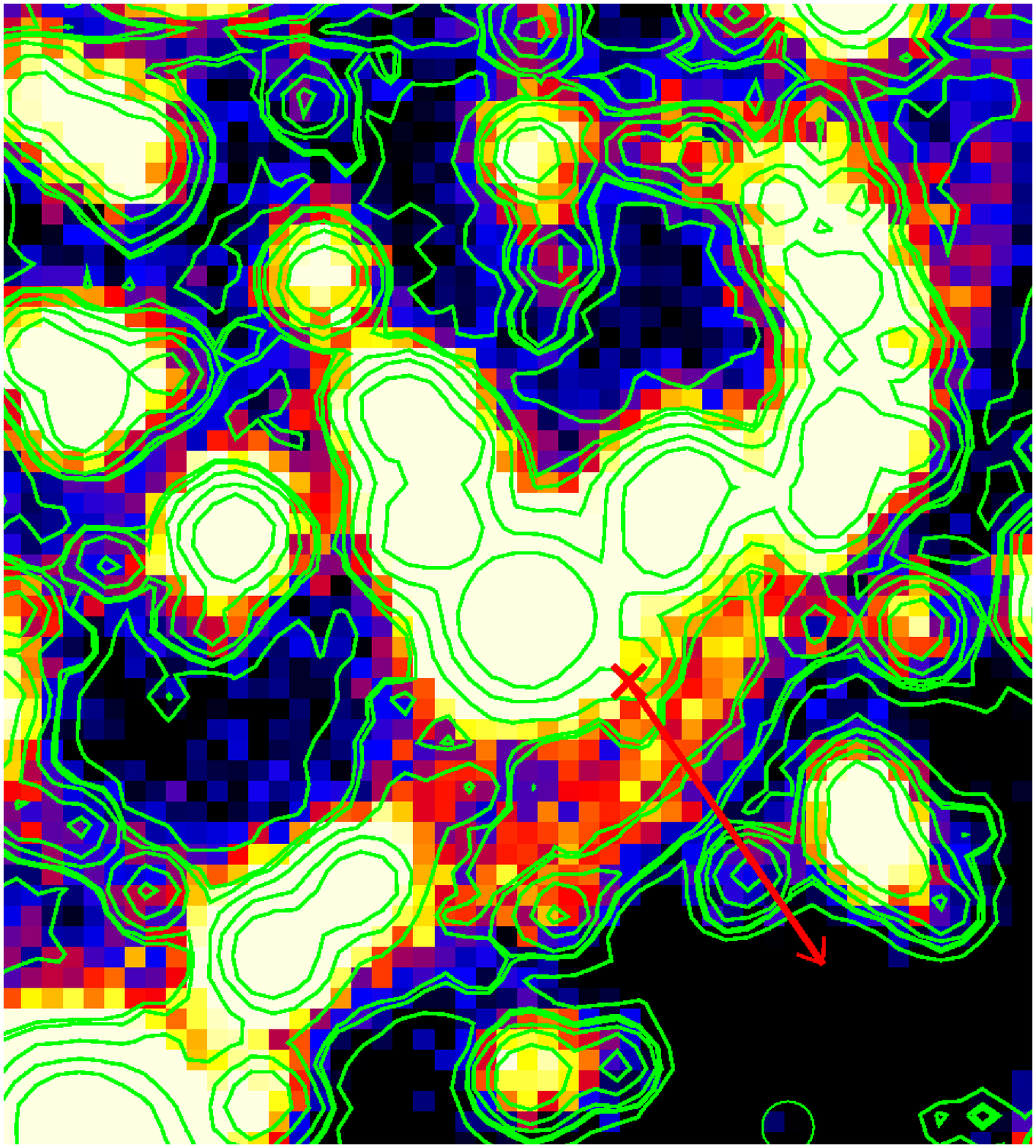}
\includegraphics[height=8cm,clip=]{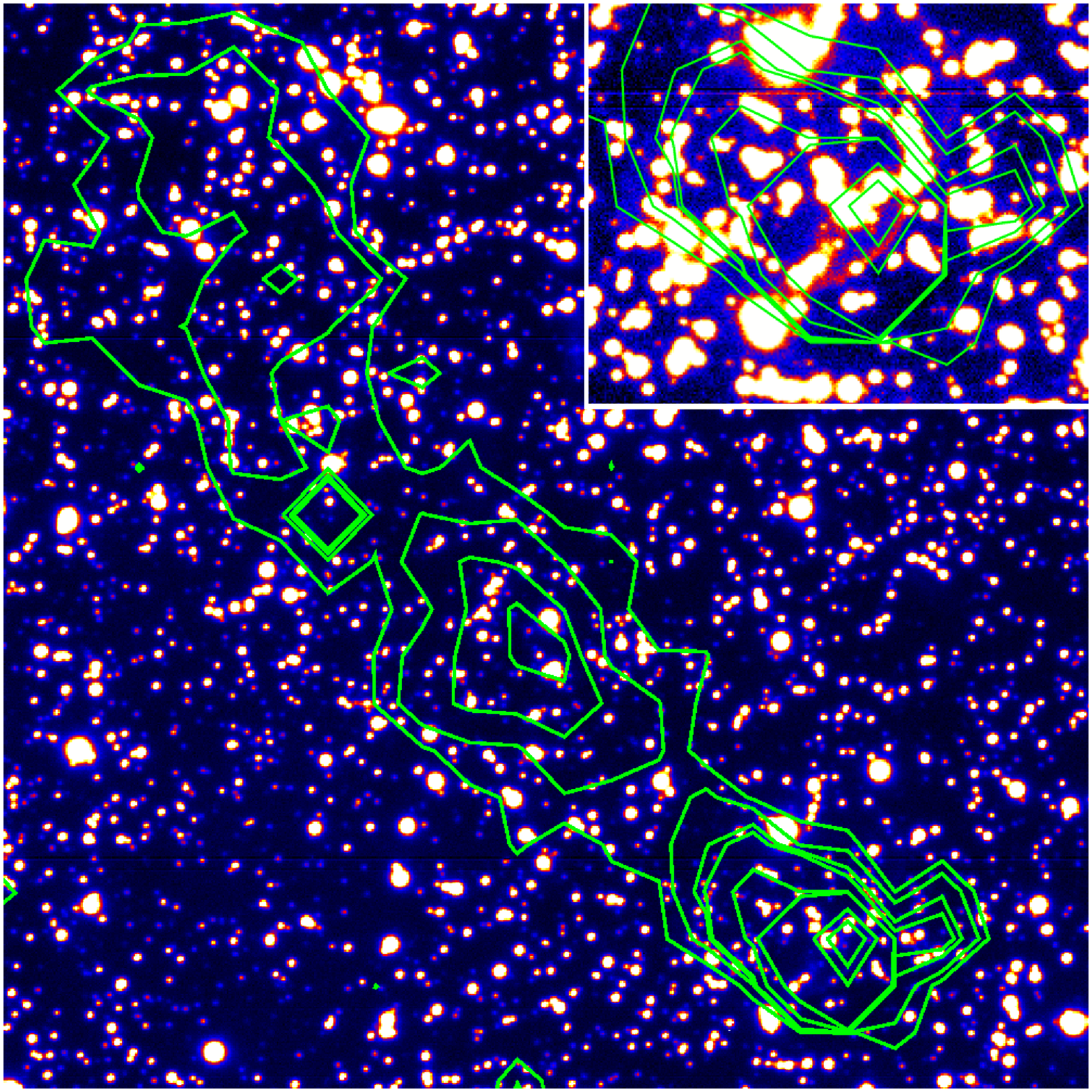}
\caption{{\em Left}: NTT H$_{\alpha}$ image of the \psr\ bow shock nebula with the contours from the FORS2 b$_{\rm HIGH}$-band image overlaid. The bow-shock displacement over the two epochs is apparent. The cross marks the pulsar position and the arrow its proper motion direction. {\em Right}: FORS2 image of the whole X-ray PWN region. The X-ray contours from the \chan\ observation of \psr\ are marked in green. The inset shows a close-up of the PWN around the pulsar position. \label{pwn}}
\end{center}
\end{figure}

On a much larger scale, we found no evidence of diffuse 
optical emission which can be associated with the X-ray PWN (Fig.\ref{pwn}, right).  From our broad-band images,  which cover the entire PWN area, we estimated $3\sigma$ upper limits of $\sim 27.8$ and $\sim 28.1$ magnitudes arcsec$^{-2}$ on its optical surface brightness in the b$_{\rm HIGH}$ and v$_{\rm HIGH}$ bands,  respectively. These limits have been computed by averaging measurements obtained in a grid of star-free regions along the PWN. 

\section{Discussion}

Owing to their faintness, only eight of the over 200 $\gamma$-ray pulsars discovered to date\footnote{\texttt{https://confluence.slac.stanford.edu/display/GLAMCOG/\\Public+List+of+LAT-Detected+Gamma-Ray+Pulsars}} have been detected at optical wavelengths (see Abdo et al.\ 2013 and references therein), the number accounting also for 
PSR\, B0540$-$69 (Caraveo et al.\ 1992), 
only recently found to be a $\gamma$-ray pulsar (Ackermann et al.\ 2015).  Furthermore, candidate counterparts have been found for PSR\, J0205+6449 (Moran et al.\ 2013) and

PSR\, J1357$-$6429 (Zyuzin et al.\ 2016).  
Here, we found a candidate counterpart to another $\gamma$-ray pulsar, \psr.
We compared its fluxes 
with the extrapolation in the optical regime of the pulsar X and $\gamma$-ray spectra.   The {\em Fermi}-LAT spectrum is described  by a PL with photon index $\Gamma_{\gamma}=1.04\pm0.07$ and an exponential cut-off at energy $E_{\rm cut} = 0.88\pm0.05$ GeV, yielding a flux $F_{\gamma}= (11.8\pm0.28) \times 10^{-11}$ erg cm$^{-2}$ s$^{-1}$ (Acero et al.\ 2015). The joint \xmm\ and \chan\ spectrum (Marelli et al.\ 2014) is described by a combination of a PL with photon index $\Gamma_{\rm X} = 2.68\pm0.04$ and a BB of temperature $k T= 0.060\pm0.0016$ keV, with a corresponding emitting area of radius $5.39^{+0.81}_{-0.71}$ km for a pulsar distance of 0.38 kpc.  The spectral fit yields an  $N_{\rm H}=(1.21\pm0.01) \times 10^{21}$ cm$^{-2}$ and unabsorbed fluxes of $F_{\rm X}^{\rm PL} =  (5.47\pm0.13)\times 10^{-13}$ and $F_{\rm X}^{\rm BB} =  (7.63\pm0.19)\times 10^{-13}$ erg cm$^{-2}$ s$^{-1}$ for the PL and BB components, respectively. Similar spectral parameters were obtained by fitting the \chan\ data alone with the same spectral model (Karpova et al.\ 2014; Auchettl et al.\  2015). We computed the interstellar reddening along the line of sight, $E(B-V)\sim0.22$, from the $N_{\rm H}$ 
and the relation of Predehl \& Schmitt (1995).  Then, we computed the unabsorbed spectral fluxes of the pulsar candidate counterpart in the b$_{\rm HIGH}$ and  v$_{\rm HIGH}$ bands applying the interstellar extinction coefficients of Fitzpatrick (1999). 

\begin{figure}
\vspace{0.5cm}
\begin{center}
\includegraphics[width=10.5cm.,clip=]{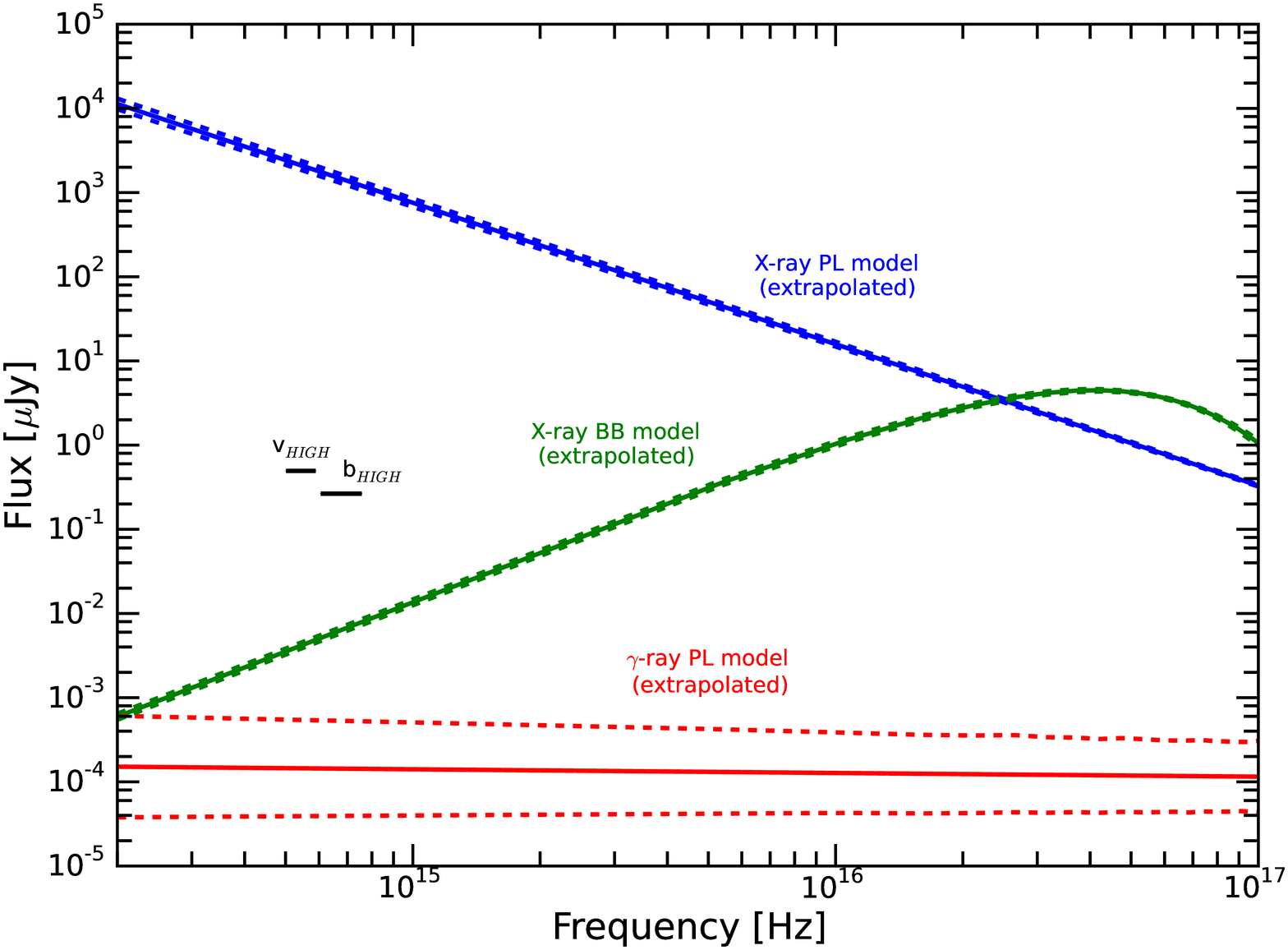}
\caption{Spectral energy distribution of \psr. The optical flux measurements are labelled with the filter names. The blue and red lines represent the extrapolation in the  optical regime of the PLs best-fitting the X and $\gamma$-ray spectra, respectively, whereas the green curve corresponds to the BB component to the X-ray spectrum. The dashed lines correspond to the $1 \sigma$ error. \label{sed}}
\end{center}
\end{figure}

The b$_{\rm HIGH}$ and  v$_{\rm HIGH}$-band optical fluxes 
lie well below the extrapolation of the X-ray PL, and above that of the $\gamma$-ray PL  (Fig.\ref{sed}),  suggesting that the non-thermal X-ray spectrum breaks at low energies. Single/multiple breaks in the multi-wavelength spectrum are observed in $\gamma$-ray pulsars, without an obvious relation to their characteristics, e.g. age and magnetic field (Mignani et al.\ 2016).  More likely, they are related to different geometries of the emission regions in the pulsar magnetosphere and different viewing angles, which produce different light curve profiles. In the case of \psr, the break between the X and $\gamma$-ray PLs  indeed comes with a difference in the X/$\gamma$-ray light curves (Marelli et al.\ 2014). 

Like in the X-rays, we expect that the optical emission of \psr\ results from both the non-thermal emission from the neutron star magnetosphere and the thermal emission from the cooling neutron star surface as observed in other middle-aged pulsars, i.e.  PSR\, B0656+14, Geminga (Kargaltsev \& Pavlov 2007), and PSR\, B1055$-$52 (Mignani et al.\ 2010). 
Indeed, the optical fluxes are above the extrapolation of the BB component to the X-ray spectrum and are not consistent with a Rayleigh-Jeans spectrum. This implies both that
any thermal component to the optical spectrum must be produced from a region on the neutron star surface presumably larger and colder than that producing the X-ray emission and that a non-thermal PL component must be present.  Obviously, with only two flux measurements it is impossible to decouple these two components.  
Multi-band follow-up observations are required for a spectral characterisation.
The unabsorbed flux of the candidate counterpart is $F_{\rm opt} \sim 4.09 \times 10^{-16}$ erg cm$^{-2}$ s$^{-1}$, integrated over the v$_{\rm HIGH}$ band. This corresponds to a luminosity $L_{\rm opt} \sim 7.06 \times 10^{27}$ erg s$^{-1}$ for a distance of 0.38 kpc. The ratio to the pulsar rotational energy loss rate yields an optical emission efficiency  $L_{\rm opt} /E_{\rm rot} \sim 7.4 \times 10^{-7}$. We also compared the unabsorbed optical flux with the unabsorbed, total X-ray flux $F_{\rm X}$ and the $\gamma$-ray flux $F_{\gamma}$.  This yields $F_{\rm opt}/F_{\rm X}\sim 3.1 \times 10^{-4}$ and $F_{\rm opt}/F_{\gamma}\sim 3.5 \times 10^{-6}$. All these values are in the range of those computed for the other middle-aged pulsars identified in the optical (see, e.g. Tab.\ 4 of Moran et al.\ 2013). This confirms that middle-aged pulsars tend to have similar multi-wavelength emission properties.

The better spatial resolution and sensitivity of our FORS2 images with respect to those of Romani et al.\ (2010) make it possible to better resolve the morphology of the bow-shock nebula (Fig.\, 2, left). Romani et al.\ (2010) noted that its shape is remarkably different with respect to the expectations of the model by Wilkin (1996), which assumes an isotropic pulsar wind. These authors computed the ratio of the perpendicular half-angular size of the nebula $\theta_{\perp}$ (measured through the pulsar) to the separation $\theta_{\parallel}$ between the pulsar and the apex of the nebula as a measure of the "flatness" of the bow-shock front and showed that the observed value can be reproduced by a model assuming an equatorially-concentrated pulsar wind, aligned pulsar spin axis and space velocity, and edge-on viewing geometry. Our images allow us to  better constrain the distance of the bow-shock apex from the pulsar. Similarly to what we did in Sectn. 3, we extracted the nebula surface brightness profile along an image strip of 0\farcs625 width, aligned with the pulsar proper motion direction (Auchettl et al.\ 2015), 
and we 
measured 
its position
by fitting a simple Gaussian function. The resulting angular separation with respect to the proper motion-corrected \chan\ position of the pulsar (MJD 57156),
$1\farcs6\pm0\farcs5$,
is consistent with the value of $\sim1\farcs5$ estimated by Romani et al.\ (2010) using the original {\em Chandra} position (MJD 55337) and the NTT H$_{\alpha}$ image (MJD 55063). 
A refined measurement of the pulsar proper motion, together with a more accurate pulsar localisation from the identification of its optical counterpart, will allow us to probe in more detail the geometry of the momentum deposition by the pulsar particle wind. 

The X-ray PWN  is characterised by a segmented structure (Fig.~\ref{pwn}, right), with three bright emission lobes, one centred on the pulsar, and the other two almost aligned along the PWN tail, at $\sim40\arcsec$ and $\sim80\arcsec$ from the pulsar. The FORS2 images show that these two lobes cannot be produced by the combined emission of bright back/foreground stars but must be intrinsic to the PWN. In case of a classical X-ray synchrotron emission, a variation of the mean interstellar density would cause the formation of such lobes, although this would imply a variation of a factor $\sim$10 on a $\sim$0.01 pc scale (for a 0.38 kpc distance). Alternatively, a spatial-dependent particle re-acceleration mechanism could be at work, as already invoked for other peculiar  PWNe around, e.g. PSR\,  J2055+2539 (Marelli et al.\ 2016) and PSR\, J1509$-$5850 (Klingler et al.\ 2016), or a different emission mechanism could be responsible for powering the X-ray PWN, like, e.g. in PSR\, J0357+3205 (Marelli et al.\ 2013).

\section{Conclusions}

Using the VLT, we found a possible candidate optical counterpart to \psr\  ($m_b = 26.13$) based upon positional coincidence with its proper motion-corrected \chan\  coordinates.  \psr\ would, then, be the third $\gamma$-ray pulsar discovered by {\em Fermi} for which a candidate optical/infrared counterpart has been found, after PSR\, J0205+6449 (Moran et al.\ 2013) and PSR\, J1357$-$6429 (Zyuzin et al.\  2016).
Multi-epoch optical observations will allow us to measure the proper motion of the \psr\ candidate counterpart and confirm the optical identification, whereas multi-band photometry will be needed to determine the pulsar spectrum in the optical.   In the same VLT data, we also detected the bow-shock nebula around the pulsar and found that it is displaced to the southwest with respect to its position measured in the 2009 NTT data of Romani et al.\ (2010). The annual displacement ($169^{+35}_{-54}$ mas yr$^{-1}$) is compatible with that expected for the pulsar proper motion, showing that the shock propagates in the ISM as the pulsar moves through it. Finally, we looked for extended optical emission associated with the X-ray PWN but we could not detect it down to a limit of $\sim 28.1$ magnitudes arcsec$^{-2}$  in the $v_{\rm HIGH}$ band, the deepest obtained so far for this nebula.

\acknowledgments
We thank the anonymous referee for his/her constructive comments to our manuscript. RPM acknowledges financial support from the project TECHE.it. CRA 1.05.06.04.01 cap 1.05.08 for the project "Studio multilunghezze d'onda da stelle di neutroni con particolare riguardo alla emissione di altissima energia". The work of MM was supported by the ASI-INAF contract I/037/12/0, art.22 L.240/2010 for the project $"$Calibrazione ed Analisi del satellite NuSTAR$"$.

{\it Facilities:} \facility{Very Large Telescope}

\end{document}